%% file: main.tex
\documentclass[manuscript]{acmart}

\AtBeginDocument{%
  \providecommand\BibTeX{{%
    \normalfont B\kern-0.5em{\scshape i\kern-0.25em b}\kern-0.8em\TeX}}}

\setcopyright{none} 

\usepackage{xspace}

\usepackage{subfigure} 

\usepackage[labelfont=bf]{caption}

\newcommand{\ucb}{\textit{Brahmins}\xspace}
\newcommand{\ofc}{\textit{Other Forward Castes}\xspace}
\newcommand{\ic}{\textit{Intermediary Castes}\xspace}
\newcommand{\obc}{\textit{Other Backward Castes}\xspace}
\newcommand{\scs}{\textit{Scheduled Castes}\xspace}
\newcommand{\st}{\textit{Scheduled Tribes}\xspace}

\newcommand{\upperc}{\textit{Upper-Caste}\xspace}
\newcommand{\lowerc}{\textit{Lower-Caste}\xspace}

\begin{document}

\title[Indian Political Twitter and Caste Discrimination]{Indian Political Twitter and Caste Discrimination - How Representation Does Not Equal Inclusion in Lok Sabha Networks}



\author{Palashi Vaghela}
\affiliation{%
  \institution{Cornell University}
  \city{New York}
  \country{USA}
}
\email{palashi@infosci.cornell.edu}

\author{Ramaravind Kommiya Mothilal}
\affiliation{%
  \institution{Microsoft Research}
  \city{Bengaluru}
  \country{India}
}
\email{t-rakom@microsoft.com}

\author{Joyojeet Pal}
\affiliation{%
  \institution{Microsoft Research}
  \city{Bengaluru}
  \country{India}
}
\email{Joyojeet.Pal@microsoft.com}

\renewcommand{\shortauthors}{}

\begin{abstract}
 Caste privilege persists in the form of upper caste ``networks'' in India made up of political, social, and economic relations that tend to actively exclude lower caste members. In this study, we examine this pernicious expression of caste in the Twitter networks of politicians from India's highest legislative body - the Lok Sabha.  We find that caste has a significant relationship with the centrality, connectivity and engagement of an MP in the Lok Sabha Twitter network. The higher the caste of a Member of the Parliament (MP) the more likely they are to be important in the network, to have reciprocal connections with other MPs, and to get retweeted by an upper caste MPs.

\end{abstract}





\maketitle

\input{introduction}
\input{methods}

\input{findings}
\input{acknowledgements}

\bibliographystyle{ACM-Reference-Format}
\bibliography{references}

\end{document}

%% file: introduction.tex
\section{Background}

The role and performance of networks are critical, especially in positions of power. One setting to observe caste-based fraternizing, where individual actors can both act and enact power, is in the parliament. We studied Twitter networks of members of an institution which represents a diversity of caste identities - the Lok Sabha. In the years since the 2014 general election, social media is increasingly the preferred means of primary communication by major political leaders - a place for them to manage their brands, communicate policy, perform politics without the mediation of professional journalists. What politicians do online - who they communicate with, follow, or retweet, are integrally tied into the ways they perform and reciprocate relationships with their networks and their electoral base. The network effects of caste relations are well known among lower caste communities who are often treated as token diversity or restricted access to important connections or promotions in workplaces. But, how exactly do these network effects result in exclusion of Dalit, Bahujan or Adivasi people?
 
We asked three questions to this network: -
1. Is there a relationship between the caste of a Lok Sabha member and how they choose to follow other Lok Sabha members? We measured this by looking at how well an MP's following is reciprocated in the network.
2. Is there a relationship between the caste of a Lok Sabha member and how important they are in the network? We measured this by looking at how well followed an MP is in the network and how parsimonious they are with their connections.
3. Is there a relationship between the caste of a Lok Sabha member and whom they choose to engage with? We measured this through the odds of an MP getting retweeted by an Upper-Caste MP in the network.

We found that caste indeed has a very significant relationship with an MP's importance, connectivity and engagement in the Lok Sabha network. The higher your caste, the more likely you are to have a reciprocal connection with Lok Sabha members. Second, the higher your caste, the more likely you are to be important in the network. Third, the higher your caste, the more likely you are to get retweeted by a higher caste Lok Sabha member.  

%% file: methods.tex
\section{Methods}

\subsection{Data}

We used a database of Indian politicians on Twitter, created using a Machine Learning classification pipeline called \textit{NivaDuck} \cite{nivaduck}, to get Twitter profiles of politicians in India and map them to the list of Members of Parliament (MPs) through partial string matching of names. We manually checked Twitter description, tweet content, account creation date, and party. At the end of this process, we found that 489 out of 542 currently sitting MPs have a Twitter presence. Out of the 489, 465 MPs had Twitter accounts that were active at the time of our study.

We used the SPINPER database \cite{liaspinper}, that tracks historical information of caste and religion of elected MPs of India, to annotate the castes of the members of Lok Sabha. Through a combination of string matching and manual data processing, we classified the 449 MPs from Hindu religion and into six caste categories: \ucb, \ofc (Non-Brahmin Upper Castes), \ic (IC), \obc (OBC), \scs (SC), and \st (ST). Intermediary castes indicate a group of castes that are historically \upperc but are fighting for a change of status to \lowerc through political and social movements, for example, Jats, Lingayaths, Patels, etc. For the purpose of this study, we are not looking at castes in religions other than Hinduism, although we acknowledge that they exist and play an active role in the political landscape of India. We took the Hindu caste category to be an ordinal variable based on the graded hierarchy as understood in the \textit{varna} system -- \ucb $>$ \ofc $>$ \ic $>$ \obc $>$ \scs $>$ \st \cite{dumont1980homo,gupta1980varna}.

The data on follow networks of Lok Sabha MPs was collected using the public API of Twitter. We considered links made between two MPs in the period of March 2019 and February 2020 owing to important socio-political events like General Elections, legislative and constitutional changes following the elections, etc. We believe this would have shifted the dynamics of who follows whom over the period of that year. We found that 308 MPs of the 465 active Twitter accounts follow each other. The follow links help us understand network properties like centrality and reciprocity. As a way of understanding relationships of engagement among the MPs in this period, we look at the odds of being retweeting in the network by an \upperc MP.  

We assumed that the political party of the MP would interplay with caste in terms of connections, centrality and engagement because of ideological differences and ascension dependent on whether you are part of the ruling or opposition. Close to 60\% of the sitting MPs belong to Bharatiya Janata Party) (BJP,) and Indian National Congress (INC) has only 14\% of BJP's strength. Thus we bifurcated the MPs as belonging to BJP or non-BJP, and this is a distinction we maintain for the rest of the paper referring to each of the groups as ``party'' to improve comprehension.

\subsection{Structural Network Measures}

Studies done in sociology and anthropology point to caste as a form of relation that approaches the idea of inclusion and exclusion as a matter of relative effect of whether or not one is part of a network. In fact, some scholars have suggested that social networks might be a more effective way to understand abstract relations between groups. One of the ways of tracing these relations is to look at their concrete manifestation \cite{vijayabaskar2014caste}. In our case, whether you decide to follow or engage with somebody working in the same institution as you, count as concrete measures of being related to someone on Twitter. In a caste society, to understand how each network relates to each other as a whole (for example a caste category determined through scriptures and kinship,) one would need to abstract and synthesize the group level effects of how these relations dynamically unfold between different caste categories \cite{srinivas1964212}.

To that end, first, we studied two structural properties of the follow network of MPs - Reciprocity and PageRank Centrality \cite{wasserman1994social,borgatti2006graph}. When observed at an aggregate level between different groups of caste, it helped us understand how caste based exclusion can manifest online in the Lok Sabha network. We compared the median values between each pair of castes within BJP and Non-BJP separately. While not exhaustive, these two properties ascertain some of the most important and subtle insights about influence of caste on following behaviour.

Further, the reciprocity and centrality measures of a node rely on  information from other nodes in the network and hence are not independent. Thus, the standard statistical assumptions about independent observations and data distributions do not hold. We resort to permutation tests to compute the significance of the difference between medians, similar to other studies that compare network properties \cite{farine2017guide,simpson2013permutation}. We randomly swapped the positions of MPs in the network 10,000 times and estimated the difference in medians between two groups in each trial. Next, we compared this distribution of differences to the observed difference in medians. The p-value equals the proportion of differences that are at least as extreme as the observed value, and we considered the difference to be significant at 0.05.

\subsection{Retweets and Odds Ratios}
Retweets indicate a wide range of functions, such as publicly validating a message or signaling an alignment of interests. \cite{boyd2010tweet,macskassy2011people,wu2011says,luo2013will,yang2010understanding,nagarajan2010qualitative}. To understand the role of party and caste hierarchy in getting higher retweets, we sought to understand if belonging to a particular caste had any association with getting retweeted by \upperc MPs. The goal of this research question was to understand if castes higher up in the ladder prefer retweeting tweets of specific castes versus others.  Empirically, for each pair of castes, we use a Fisher's exact test to determine the significance of relative odds of getting retweeted by upper castes.

%% file: findings.tex
\section{Findings}

We find that while the relationships between caste show important patterns across party, the nature of the parties and their network structure impacts who connects with, reciprocates, and amplifies the message of which parliamentary colleague.

\begin{figure*}
  \centering
    \subfigure[Reciprocity\textsubscript{caste1} - Reciprocity\textsubscript{caste2} \label{reciprocity}]{\includegraphics[width=\columnwidth]{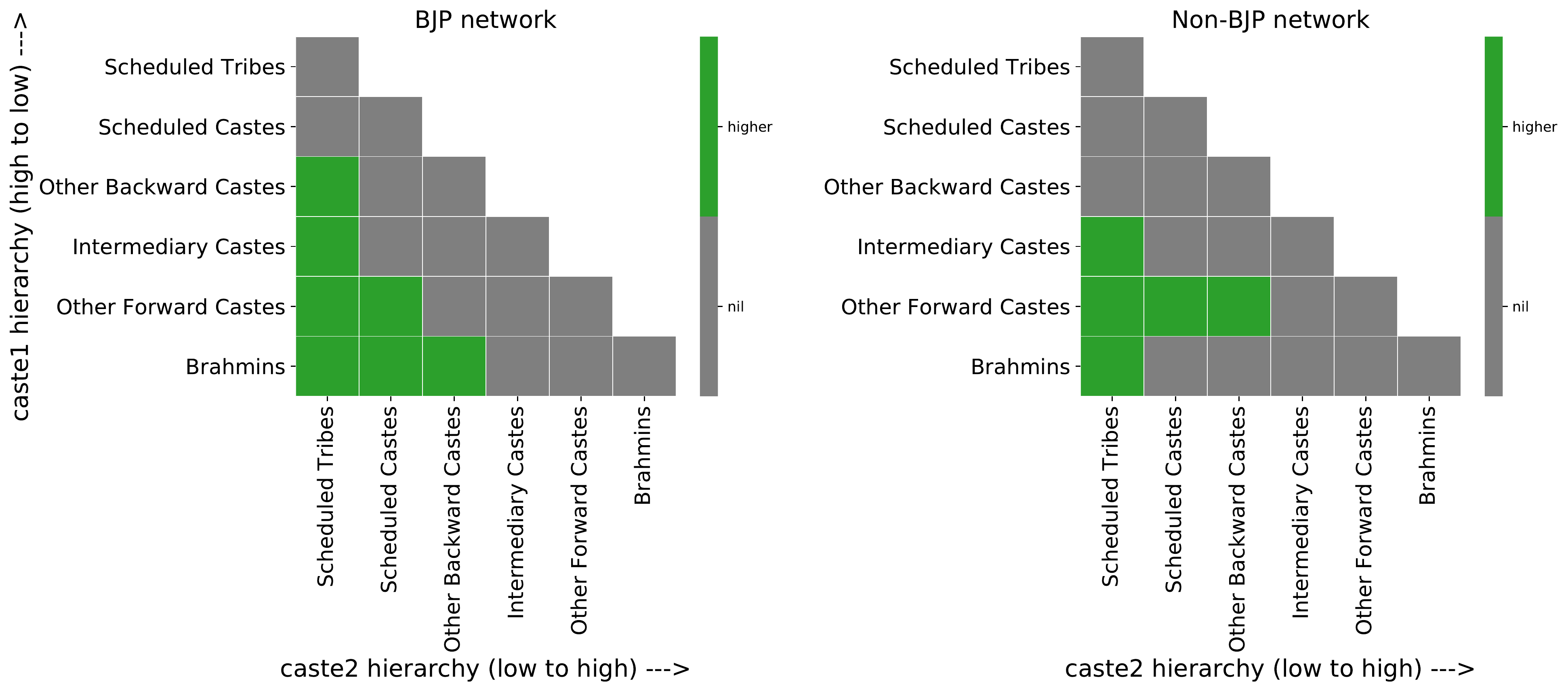}} \vspace{0.5cm}
    
    \subfigure[PageRank\textsubscript{caste1} - PageRank\textsubscript{caste2}\label{pagerank}]{\includegraphics[width=\columnwidth]{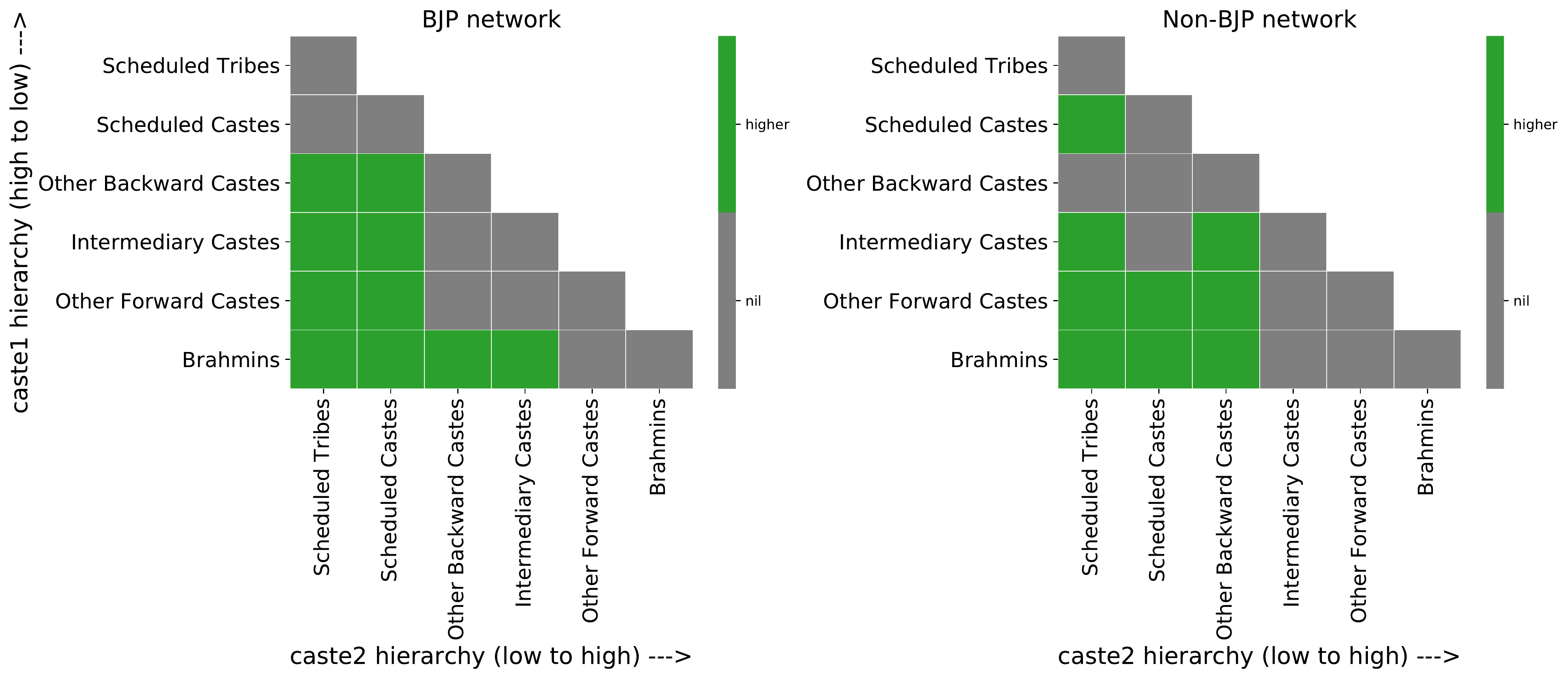}} \vspace{0.5cm}
    
  \caption{The left panes in each of the figures (a,b) show the difference between the median value of each network property, Reciprocity and PageRank centrality between each pair of castes within BJP MPs. The right panes show the same within Non-BJP MPs. The color green or red indicate if caste 1 (rows) has a significantly higher or lower median than caste 2 (columns) respectively. The grey cells correspond to insignificant differences. The significance is computed using permutation tests and all non-zero differences are significant at 0.05 level.}
  \label{median-heatmap}
\end{figure*}

\subsection{Reciprocity and Party Affiliation}

\noindent \textbf{Within the BJP:} We see that \ucb MPs have a significant advantage over all the \lowerc MPs in terms of getting reciprocal \textit{following} links from other MPs in the network in Figure \ref{reciprocity}. As we find in the bottom-most row, their median reciprocity is significantly higher than that of all \lowerc categories. 

This advantage also spreads to other upper caste MPs. As we see in second row from the bottom, \ofc get more reciprocal links from \scs and \st, and this is then followed by \ic who have higher median reciprocity than \st. 

In summary, \ucb seem to dominate in terms of forming reciprocal links in the BJP, while \st MPs have the least proportion of mutual connections to the number of MPs they follow. Essentially, a strong set of conscious or unconscious relationships of reciprocity are visible in these connections.
\\
\\
\noindent \textbf{Among non-BJP parties:} We felt that the relationships between MPs may be more complicated because of cross party engagements and may confound the effect of caste on network properties, but we also see some comparables to the patterns seen with the BJP.

One difference is seen in the right panel of Figure \ref{reciprocity}. We find that \st MPs are less reciprocated in their ``following" than all \upperc MP. Also, while  \ofc MPs, form only a small share of the sample at about 10\% of Non-BJP MPs, they seem to have higher reciprocity than all lower caste MPs from non-BJP parties, including the \obc who are more than 30\% of the total strength. 

There is however one important symmetrical property that appears to exist across parties. The caste categories that have highest median reciprocities, \ucb in BJP and \ofc in Non-BJP, have significantly less variance in reciprocity than the castes with next higher median values.    

The findings here could imply that the tendency among \upperc MPs to follow back those who send them following requests is low, or conversely, that \lowerc have a lower likelihood of getting followed back by others. 

It is clear that the propensity or ability to form mutual links by \upperc MPs is significantly greater than \lowerc, indicating that perhaps the social capital and other factors affecting \lowerc MPs are shaping how they can effectively be excluded by \upperc MPs in the network. 

Since we cannot tell the temporal relationships -- i.e. who followed whom and when, one way to think of these findings is that if an \upperc MP chooses to publicly ``follow'' or in other words ``listen" to someone in the network, they are more likely to be heard in return than \lowerc MPs.  This in turn implies that lower caste MPs are not heard as widely in the network as their counterparts from upper castes, and raises the related question: What is the position that an MP holds in the network in terms of their connectedness and importance, and how can this be related to engagement, and what in turn does this have to do with caste identity?

\subsection{PageRank Centrality}

\noindent \textbf{Within the BJP:} Similar to the trends observed in reciprocity, Figure \ref{pagerank} (left) shows that \ucb have higher median PageRank centrality, thereby hold structural positions in the network that are advantageous in exerting greater influence in information diffusion than \lowerc MPs. In fact, \ucb also have higher median PageRank than \ic. 

We observed that the \ucb are almost similar in numbers compared to most other castes in the Lok Sabha. However, they get followed by a relatively higher number of MPs, including other most followed MPs in the network, thus scoring high median PageRank centrality. On the other hand, \scs and \st are placed at a significant disadvantage in the network where they have limited in-network following, thus having least control over how far the information they broadcast gets consumed in the networks of power. 

Further, \ofc also have significantly higher variance in PageRank than any other castes. These patterns are thus similar to what was found previously in the context of forming reciprocal links.
\\
\\
\noindent \textbf{Among non-BJP parties:} Figure \ref{pagerank} (right) shows that both \ucb and \ofc have higher median PageRank within the Non-BJP networks. While this is in alignment with the \upperc dominance we have seen so far, however, \scs have a higher median pagerank than \st MPs. Similarly, \upperc MPs have higher variance than \st MPs alone, making \st the worse-off caste within Non-BJP. 

The above findings highlight that, within BJP, occupying influential positions in the network is associated with higher \textit{following} reciprocity, and we indeed find that they are well correlated with a Pearson correlation of 0.8 (p-value$<$0.0001).  

However, the correlation is relatively weaker within Non-BJP (Pearson correlation of 0.5, p-value$<$0.0001). Here, we see that \ucb, who have higher median PageRank over all the \lowerc MPs, do not have higher Reciprocity. Similarly, while \scs within Non-BJP have lower odds of reciprocal links, they occupy relatively important positions compared to \st and thus have higher median PageRank. 

In summary, both Reciprocity and PageRank together show that \upperc MPs indeed have a greater ease or ability to form reciprocal connections and are most likely to have higher control over the information consumption and broadcasting. Similar to the median comparisons, we conducted all correlation tests using a permutation-based approach.

\subsection{Retweets odds ratio and caste hierarchy} 

\begin{figure}[ht]
\centering
  \includegraphics[width=\columnwidth]{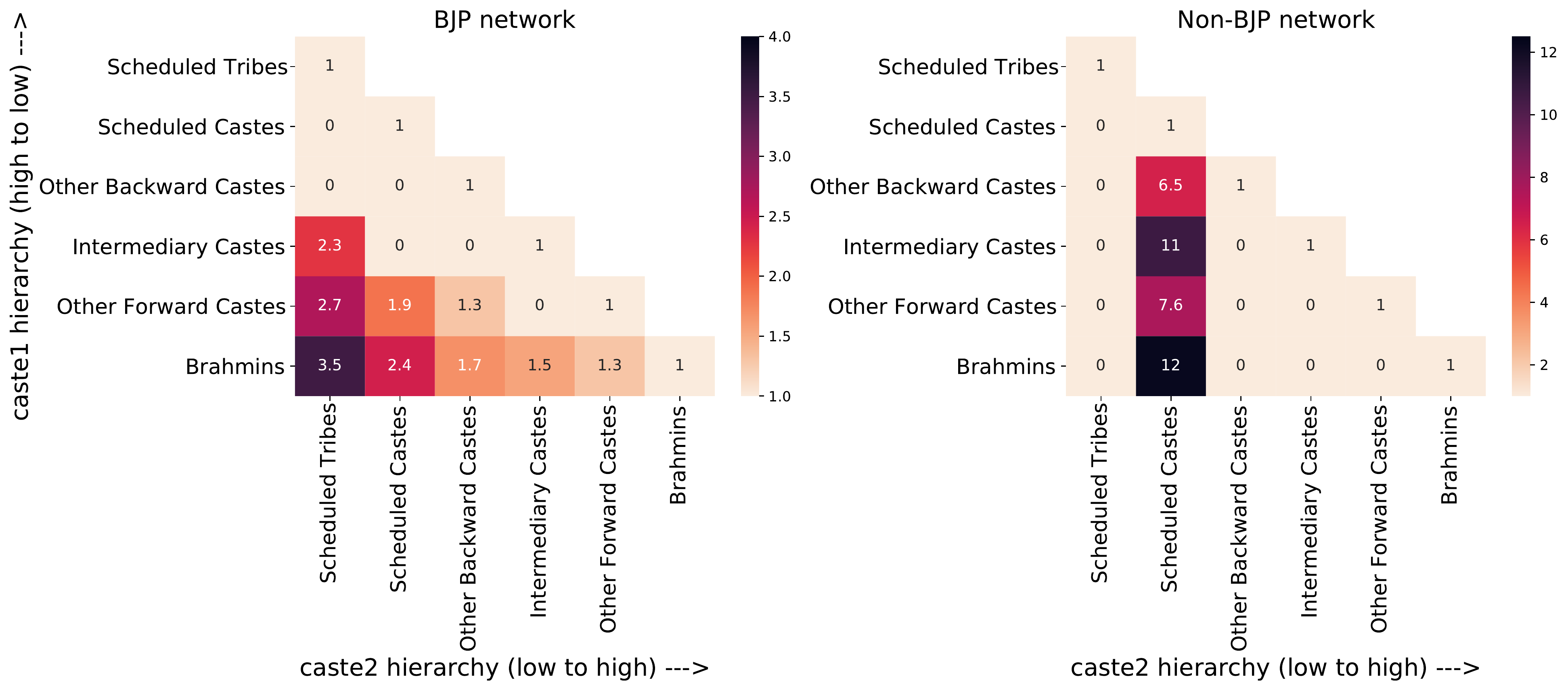}
  \caption{\textbf{The left figure shows the ratios of odds of being retweeted by \upperc MPs between each pair of castes \textit{within BJP}, and the right figure shows the odds ratios (ORs) \textit{within Non-BJP}. An OR for a pair of castes is computed as the ratio of odds for the caste along the row (caste1 in figure) to odds for the caste along the column (caste2 in figure). All non-zero ORs in the figure are significant at 0.05 level. The ORs at cells $a_{ij}$ and $a_{ji}$ are reciprocal of each other due to symmetry. }}~\label{fig:odds_uppervslower}
\end{figure}

Figure \ref{fig:odds_uppervslower} shows the pairwise Odds Ratios (ORs) of getting retweeted by \upperc MPs determined using Fisher's exact test. It can be observed that ORs are higher (darker regions) in the lower left portion the figures, which indicates the dominance of \upperc MPs. 
\\
\\
\noindent \textbf{Within the BJP:} The bottom-most row in figure \ref{fig:odds_uppervslower} (left) shows that, within BJP, the odds of being retweeted by \upperc MPs is significantly higher for \textit{Brahmin} MPs than those of any other caste. As we we move right from the left-most cell in the left figure, as we go further up the caste hierarchy, we see odds of getting retweeted by \upperc improving in comparison to that of a relatively lower caste. For example, the odds of \textit{Brahmin} MPs getting retweeted by \upperc MPs is 1.5 times higher than the odds of \textit{Intermediary Caste} MPs being retweeted by the same \upperc network. Yet, this is much better than the odds of \st or \scs MPs who are 3.5 times and 2.4 times less likely to be retweeted than \ucb MPs.

Similarly moving up a row from \ucb, the odds of being retweeted as a \ofc MPs is significantly higher for \textit{Scheduled Caste} and \textit{Scheduled Tribe} MPs at 1.9 and 2.7 respectively. Overall, they are doing much better than all \lowerc. Next is \ic, in the row above, who seem to be at an advantage of 2.3 times better odds of being retweeted by \upperc over arguably the lowest caste in the pool - \st. 

The darker regions in the bottom left cells of the Figure \ref{fig:odds_uppervslower} (left) thus indicate that \lowerc's odds of getting retweeted in the upper caste network is pretty low within BJP. Particularly, the numbers in the first column of the figure shows that \textit{Scheduled Tribe} MPs have the lowest odds of getting retweeted in the \upperc network. 

Further, since the plot is symmetrical along the diagonal, we also interpret that all \lowerc MPs, especially, \textit{Scheduled Tribe}, have higher odds of being retweeted by lower caste MPs than upper caste MPs. In other words, our results indicate the presence of a \textit{homophilic} relationship among the networks of \upperc and \lowerc MPs. 
\\
\\
\textbf{Among Non-BJP parties.} Similar to the trends observed in BJP, the \ucb dominate over \scs within Non-BJP parties too where they have higher odds of being retweeted by \upperc. However there is no significance difference in odds compared to other \lowerc MPs. In particular, \textit{Brahmins} and \textit{Intermediary Caste} MPs have odds more than 10 of being retweeted by \upperc MPs against \scs. 

Thus in contrast to a relatively strong homophily trend observed within BJP, for Non-BJP, the polarization of \upperc and \lowerc is relatively weaker, with the exception of \textit{Scheduled Caste} MPs. 

%% file: acknowledgements.tex
\section{Acknowledgements}

We would like to thank Mohit Kumar, Gilles Verniers and Basim Nissa of the SPINPER project for collaborating with us on this research by sharing their database of sociological information on Members of LokSabha. They were kindly available to answer any doubts we had with the data, as well as explain the process of data collection in detail. We would also like to thank Dibyendu Mishra, Anmol Panda and Zainab Akbar at Microsoft Research, India, for helping us with the process of data extraction and cleaning in the initial stages of this research.